\documentclass[conf]{new-aiaa}
\usepackage{graphicx}
\usepackage{amsmath}
\usepackage{xcolor}
\usepackage[normalem]{ulem}
\usepackage{float}
\usepackage[version=4]{mhchem}
\usepackage{siunitx}
\usepackage{longtable,tabularx}
\usepackage{booktabs}
\usepackage{multirow}
\usepackage{array}
\usepackage{url}
\title{Efficient Projection-Based Reduced-Order Modeling for Computational Fluid Dynamics via Multifidelity Basis Construction}

\author{Ashok Viswanathan\footnote{Ph.D. Student, Department of Aeronautics and Astronautics, Stanford University, Stanford, CA 94305, USA.}}
\affil{Stanford University, Stanford, CA 94305, USA}

\author{Charbel Farhat\footnote{Vivian Church Hoff Professor of Aircraft Structures, Department of Aeronautics and Astronautics, Stanford University, Stanford, CA 94305, USA.}}
\affil{Stanford University, Stanford, CA 94305, USA}

\begin{document}

\maketitle

\begin{abstract}
Projection-based reduced-order models (PROMs) can substantially reduce the cost of repeated computational fluid dynamics (CFD) 
simulations, but constructing an accurate reduced-order basis (ROB) remains expensive when each high-fidelity (HF) snapshot requires a
Reynolds-averaged Navier--Stokes (RANS) or higher-fidelity simulation. This work presents a multifidelity methodology that reduces the 
offline cost of projection-based model order reduction (PMOR) by combining a limited number of HF snapshots with more extensively 
sampled low-fidelity (LF) solutions. The methodology accounts for two distinct sources of fidelity discrepancy: the governing physical
model and the spatial discretization. HF and LF solutions are first computed and aligned at a small set of parameter points, and their
discrepancies are used to construct a correction ROB. At additional parameter points, this basis transforms LF predictions into 
approximations of HF solutions by minimizing the HF residual over an affine trial space, with residual-based RANS smoothing applied 
when necessary. The accepted snapshots adaptively enrich the training database and update the correction ROB, so that fully converged 
HF simulations are required only for initialization. The resulting ROB can be used with conventional PMOR formulations, including 
least-squares Petrov–Galerkin (LSPG) projection. The methodology is demonstrated for steady-state transonic flow over a NACA0012 
airfoil, with a fine-grid RANS model as the HF model and a coarse-grid Euler model as the LF model. With four directly computed HF 
snapshots and 36 corrected and, when necessary, smoothed LF-derived snapshots, the multifidelity procedure reduces the offline 
snapshot-generation cost by a factor of 7.8 relative to HF-only construction. At a ROB dimension of 40, over a test set emphasizing the
shock-containing region of the parameter domain, the resulting LSPG PROM predicts lift, drag, and lift-to-drag ratio with mean errors,
relative to fully converged HF solutions, of approximately 1.1\%, 3.6\%, and 3.9\%, and the streamwise skin-friction coefficient with 
a mean relative error of approximately 5\%; the corresponding errors of a PROM constructed entirely from HF snapshots are 0.18\%, 
2.9\%, 2.8\%, and 4\%. Increasing the number of directly computed HF snapshots raises the offline cost without monotonically improving 
accuracy, revealing an accuracy--cost tradeoff. These results illustrate the potential of physics-based multifidelity data to reduce 
the cost of constructing accurate PROMs for CFD while retaining compatibility with established PMOR methods.
\end{abstract}


\section{Introduction}
\label{sec:IN}

Projection-based model order reduction (PMOR) reduces the computational cost of repeatedly solving parameterized partial differential 
equations by projecting a high-dimensional discretized model onto a low-dimensional orthonormal reduced-order basis (ROB). The 
resulting projection-based reduced-order model (PROM) replaces repeated high-fidelity (HF) simulations with substantially smaller 
reduced-order problems while preserving the dominant features of the physical response. This capability is particularly valuable in 
computational fluid dynamics (CFD), where HF simulations may involve millions of degrees of freedom (dofs) and must be repeated over 
multidimensional parameter domains -- for example, for design, optimization, uncertainty quantification, control, and digital-twin 
applications. Established PMOR formulations, including Galerkin and least-squares Petrov--Galerkin (LSPG) projection, provide a 
general framework for constructing and deploying PROMs.

The effectiveness of a PROM depends on the ROB's ability to represent the solution manifold over the parameter domain of interest. 
Because constructing an accurate ROB typically requires snapshots from the underlying high-dimensional model (HDM), the cost of 
generating the snapshot database can become a computational bottleneck. Residual-based greedy sampling~\cite{binev2011convergence} 
reduces this cost by adaptively identifying parameter points at which additional snapshots are most informative, and adaptive variants
further tune the basis size at each greedy iteration using efficient error indicators~\cite{chellappa2020adaptive}. However, these 
greedy procedures still require an HF simulation at every newly selected parameter point. For CFD problems governed by 
Reynolds-averaged Navier--Stokes (RANS) equations or higher-fidelity flow models, even a modest number of HF simulations can be 
computationally expensive. Reducing the number of directly computed HF snapshots is therefore an important complement to
accelerating online PROM evaluations.

Multifidelity modeling offers a means of reducing this offline cost by combining information from computational models of different 
fidelities. Existing approaches include multilevel proper orthogonal decomposition (POD) algorithms~\cite{behzad2018multilevel}, 
multifidelity reduced-order surrogate modeling~\cite{conti2024multi}, and nonintrusive bi-fidelity 
approximation~\cite{hampton2018practical}. Multifidelity error estimation has also been used to accelerate greedy model reduction by 
replacing the costly high-fidelity error estimator~\cite{feng2023multi}, although the training snapshots themselves remain 
high-fidelity in that setting. More recently, a multifidelity POD formulation~\cite{aretz2026multifidelity} has been proposed to 
exploit extensively sampled low-fidelity (LF) data together with a limited number of HF snapshots. Exploiting multifidelity 
information in CFD, however, requires accounting for the distinct sources of fidelity discrepancy that arise in practical flow 
simulations.

Fidelity discrepancies can originate from the governing physical model or the spatial discretization. For example, an inviscid Euler 
model neglects the viscous and turbulence effects represented in a RANS model, whereas discretization fidelity reflects the spatial 
resolution used to solve the governing equations. These two sources can be varied independently or combined: an LF Euler model on a 
coarse grid and an HF RANS model on a fine grid differ in both physical modeling and spatial resolution. A multifidelity ROB 
construction strategy should therefore accommodate discrepancies arising from either source or from both simultaneously.

This work presents a physics-based, interpretable multifidelity methodology for constructing an accurate ROB for an HF PROM while 
reducing the number of directly computed HF snapshots. Unlike the foundational multifidelity approach introduced 
in~\cite{kennedy2000predicting}, which learns relationships between HF and LF outputs for scalar or low-dimensional quantities of 
interest and can be implemented using surrogate models such as Gaussian processes or neural networks, the proposed methodology 
operates directly on high-dimensional state vectors within the discrete governing-equation framework. It exploits the HF residual both
to guide parameter selection and to correct LF predictions through a dedicated correction ROB. The methodology complements rather than
replaces conventional PMOR formulations: the resulting ROB can be used with established projection methods, such as 
LSPG~\cite{carlberg2011efficient,carlberg2013gnat}.

The methodology comprises an initialization stage and an adaptive enrichment stage. During initialization, residual-based greedy 
sampling selects a limited set of training parameters for which HF and LF snapshots are computed and aligned. Their discrepancies 
define a correction ROB. During enrichment, an aligned LF prediction at each newly selected parameter serves as a reference state for 
HF-residual-based correction. The corrected state is assessed using the HF residual and, when necessary, refined through residual-based
HF smoothing. Accepted snapshots enrich the PROM training database and update the correction ROB.

The methodology is demonstrated for steady-state transonic flow over a NACA0012 airfoil, using a fine-grid RANS model for the HF 
simulations and a coarser-grid Euler model for the LF simulations. The two models differ in both physical formulation and spatial 
discretization, requiring the alignment of solutions defined on nonmatching grids. The resulting PROM is assessed in terms of its 
predictions of aerodynamic performance and skin-friction coefficients. The study also examines the effect of the number of 
HF snapshots on prediction accuracy and evaluates the computational costs of LF-HDM evaluations, multifidelity correction, and HF 
smoothing to quantify the resulting speedup and accuracy--cost tradeoff.

The methodology is formulated here for steady-state problems, for which the number of collected snapshots is often relatively small, 
making it practical to retain the full rank of the singular value decomposition (SVD) used to construct the correction ROB. No 
truncation is therefore applied, allowing the multifidelity construction procedure to be presented independently of the additional 
considerations involved in selecting a truncation threshold. The same construction principles extend to unsteady problems, although 
the larger snapshot databases typically encountered in such applications may necessitate SVD truncation.

The remainder of this paper is organized as follows. Section~\ref{sec:MS} introduces the parametric aerodynamic problem, describes the
two sources of model fidelity, and formulates the ROB construction objective. Section~\ref{sec:MRBCM} presents the multifidelity ROB 
construction methodology. Section~\ref{sec:NR} evaluates prediction accuracy, HF snapshot requirements, computational cost, speedup, 
and accuracy--cost tradeoffs for the NACA0012 problem. Finally, Section~\ref{sec:C} presents the conclusions.


\section{Multifidelity Setting}
\label{sec:MS}

\subsection{Parametric Aerodynamic Problem}

This work considers parametric aerodynamic problems in which the parameters represent flight conditions, such as freestream Mach 
number $M_\infty$ and angle of attack $\alpha$, or geometric shape parameters. The parameters and their prescribed bounds define the 
parameter domain $\mathcal{D}\subset\mathbb{R}^d$, where $d$ is the dimension of the parameter space. For any parameter vector 
$\boldsymbol{\mu} \in \mathcal{D}$, the governing steady HF model is expressed in discrete residual form as
\begin{equation*}
	\mathbf{R}_{\text{HF}}(\mathbf{s}_{\text{HF}}(\boldsymbol{\mu}); \boldsymbol{\mu}) = \mathbf{0}
\end{equation*}
where $\mathbf{s}_{\mathrm{HF}}\in\mathbb{R}^{N_{\mathrm{HF}}}$ is the discretized HF state, $N_{\mathrm{HF}}$ is the number of HF 
dofs, and $\mathbf{R}_{\mathrm{HF}}$ is the discrete HF residual operator.

\subsection{Fidelity in Physics and Discretization}

Fidelity discrepancies in CFD arise fundamentally from two distinct sources: the governing physical model and the spatial 
discretization resolution. These sources can be varied independently or in combination within the proposed PMOR framework.

\subsubsection{Physics Fidelity}

Physics fidelity refers to the governing equations and their underlying modeling assumptions. In principle, the framework accommodates
a hierarchy of flow models, ranging from linearized potential-flow models to the Euler and Navier--Stokes equations with different 
turbulence closures.

For clarity, the present study considers an HF model governed by the RANS equations with Spalart–Allmaras turbulence closure and an 
LF model governed by the inviscid Euler equations. The LF model neglects viscous boundary layers, skin friction, and turbulence 
transport, yielding a physics-based model discrepancy relative to the HF RANS formulation.

\subsubsection{Grid-Resolution Fidelity}

Grid-resolution fidelity refers to the spatial resolution used to discretize a given physical model. The framework accommodates grids 
of different resolutions and multi-element discretizations, whether generated with or without adaptive mesh refinement. Element types 
may affect the choice of grid-transfer operator used to map LF solutions onto the HF grid, but do not alter the fundamental algebraic 
structure of the method. The framework also accommodates spatial hierarchies with more than two fidelity levels.

When the LF and HF models are discretized on coarse and fine grids, respectively, with $N_{\mathrm{LF}}\ll N_{\mathrm{HF}}$, their 
snapshots must be aligned before their discrepancies can be evaluated. The grid mismatch is addressed by transferring each LF 
solution onto the HF computational grid. This alignment requires a two-step transformation:
\begin{enumerate}
	\item \textit{Spatial Interpolation}: Mapping the LF flow state onto the fine HF spatial grid via a spatial transfer operator 
		$\mathcal{I}_{\text{LF}\to\text{HF}}$:
		\begin{equation} 
			\mathbf{s}_{\text{LF}\to\text{HF}}(\boldsymbol{\mu}) = \mathcal{I}_{\text{LF}\to\text{HF}}
			(\mathbf{s}_{\text{LF}}(\boldsymbol{\mu})) 
			\label{eq:LFtoHF}
		\end{equation}
	\item \textit{State Augmentation}: Augmenting the interpolated state to match the state-vector structure of the HF model. For a 
		variable present in the HF state but absent in the LF state (such as the Spalart--Allmaras turbulence variable 
		$\tilde{\nu}$), zero padding is applied:
		\begin{equation} 
			\tilde{\mathbf{s}}_{\text{LF}}(\boldsymbol{\mu}) = \begin{bmatrix} 
				\mathbf{s}_{\text{LF}\to\text{HF}}(\boldsymbol{\mu}) \\ \mathbf{0}
				\label{eq:stildeLF}
			\end{bmatrix}
		\end{equation}
		where $\mathbf{0}$ represents a zero vector corresponding to the missing dofs. The interpretation of this zero block 
		is discussed in Section~\ref{sec:state_augmentation}.
\end{enumerate}

More generally, define the state-alignment operator $\mathcal{A}_{\mathrm{LF}\rightarrow\mathrm{HF}}$ to combine spatial interpolation
and state augmentation. The aligned LF snapshot is then
\begin{equation} 
	\tilde{\mathbf{s}}_{\text{LF}}(\boldsymbol{\mu}) = \mathcal{A}_{\text{LF}\to\text{HF}}
	(\mathbf{s}_{\text{LF}}(\boldsymbol{\mu}))
	\label{eq:stildeLF2}
\end{equation}

\subsection{Multifidelity Reduced-Order Basis Construction Objective}

The objective is to construct a ROB that approximates the HF solution manifold over $\mathcal{D}$ while reducing the cost of generating
the training database. The methodology uses two distinct ROBs. The correction ROB represents discrepancies between parameter-matched 
HF snapshots and aligned LF snapshots, while the final PROM ROB represents the solution states used to construct the PROM.

In conventional HF-only construction, every additional training parameter requires a new HF simulation. The proposed approach instead 
combines a limited set of directly computed HF snapshots with additional snapshots obtained from LF predictions through 
HF-residual-based correction and, when needed, smoothing. Let $S_{\mathrm{HF}}$ denote the database of directly computed HF snapshots 
and let $S_{\mathrm{LF,corrected}}$ denote the database of accepted corrected LF snapshots. The final PROM ROB is constructed from the
combined database

\begin{equation*}
\mathbf{S}_{\mathrm{PROM}}= \left[ \mathbf{S}_{\mathrm{HF}}, \mathbf{S}_{\mathrm{LF,corrected}} \right]
\end{equation*}

All columns of this database are expressed on the HF computational grid and in the HF state representation. Consequently, the final 
PROM ROB approximates the HF solution manifold, whereas the correction ROB represents the space of HF–LF discrepancies used to 
generate additional training snapshots.

The method does not eliminate HF computations. Instead, it uses a limited set of directly computed HF snapshots for initialization and
subsequently exploits the HF residual and solver to correct and refine LF-derived snapshots. The intended computational benefit is 
obtained when the combined cost of LF evaluation, reduced-order correction, and HF smoothing is smaller than the cost of generating 
the corresponding fully converged HF snapshot through a conventional simulation.

The construction therefore involves an accuracy--cost tradeoff. Corrected LF snapshots are accepted according to prescribed HF-residual
criteria, but need not satisfy the same convergence tolerance as directly computed HF snapshots. Accordingly, the quality of the 
resulting ROB must be assessed through the accuracy of the associated PROM at parameters not used during training, together with the 
total computational cost of basis construction.

The proposed contribution is an approach to generating and adaptively enriching the training database and its ROB, rather than a new 
projection formulation. Once constructed, the ROB can be supplied to a conventional PMOR method for subsequent HF reduced-order 
prediction.


\section{Multifidelity Reduced-Order Basis Construction Methodology}
\label{sec:MRBCM}

\subsection{Overview of the Computational Framework}

The computational framework comprises an initialization stage and an adaptive multifidelity enrichment stage. It maintains two 
distinct ROBs: a correction ROB that represents HF–LF discrepancies and a final PROM ROB constructed from the combined training 
database.

During initialization, a standard residual-based HF greedy procedure selects a limited set of training parameters and generates the 
corresponding HF snapshots. The LF model is then evaluated at the same parameter points. The resulting parameter-matched snapshots 
are aligned through spatial interpolation and, when necessary, state augmentation. Their differences define a discrepancy matrix from 
which the correction ROB is constructed. The initial PROM ROB is constructed from the directly computed HF snapshots.

During adaptive enrichment, the HF residual of the current PROM prediction is evaluated over a candidate parameter set to select the 
next training parameter. At this parameter, the LF model is evaluated instead of generating a new HF snapshot through a conventional 
full-order simulation. The LF prediction is aligned with the HF state representation and used as a parameter-dependent shift vector. 
A correction is then determined within the correction ROB by minimizing the HF residual.

The corrected state is assessed using the HF residual and is accepted directly if its residual is lower than that of the current PROM 
prediction. Otherwise, HF RANS smoothing is applied to reduce the residual before the state is reconsidered for enrichment.

Each accepted enrichment updates both ROBs. The PROM ROB is enriched with the accepted corrected LF-derived state, while the 
correction ROB is updated with the discrepancy between that state and its parameter-matched aligned LF prediction.

This sequence is repeated until the prescribed greedy stopping criterion is met or the training budget is exhausted. The resulting 
final ROB is then used with the selected PMOR formulation for subsequent reduced-order evaluations. The following subsections describe
each component of this construction.

\subsection{High-Fidelity Sampling and Corresponding Low-Fidelity Evaluations}

\subsubsection{Residual-Based Greedy Sampling}
\label{sec:greedy}

The HF training parameters are selected using a residual-based greedy procedure. Starting from an initial
set of parameters, HF simulations are performed and the resulting snapshots are used to construct an initial PROM.

Once $k$ training parameters have been sampled, forming the set $\mathcal{P}_{\mathrm{HF}}^{(k)}$, the current PROM is evaluated over 
the candidate set $\mathcal{D}_{\mathrm{cand}} \subseteq \mathcal{D}$. The norm of the HF residual of each reconstructed PROM solution defines the sampling indicator:
\begin{equation*} 
	\eta_k(\boldsymbol{\mu}) = \left\| \mathbf{R}_{\mathrm{HF}} \left( \mathbf{s}_{\mathrm{PROM}}^{(k)}(\boldsymbol{\mu}); 
	\boldsymbol{\mu} \right) \right\|_2
\end{equation*}
where $\mathbf{s}_{\mathrm{PROM}}^{(k)}$ is the solution reconstructed by the current ROB.

The next training parameter is selected as
\begin{equation*}
\boldsymbol{\mu}_{k+1} = \underset{ \boldsymbol{\mu}\in \mathcal{D}_{\mathrm{cand}} \setminus \mathcal{P}_{\mathrm{HF}}^{(k)} }
	{\arg\max} \; \eta_k(\boldsymbol{\mu})
\end{equation*}
where $\mathcal{P}_{\mathrm{HF}}^{(k)}$ denotes the set of $k$ parameters already sampled. An HF simulation is then performed at 
$\boldsymbol{\mu}_{k+1}$, and its snapshot is added to the training database and used to enrich the ROB. 

This procedure is repeated until a prescribed stopping criterion is satisfied, such as a residual-indicator tolerance or a maximum 
sampling budget. The resulting HF training parameter set is denoted by
\begin{equation*}
\mathcal{P}_{\mathrm{HF}} = \left\{ \boldsymbol{\mu}_1,\ldots,\boldsymbol{\mu}_{N_s} \right\}
\end{equation*}
During this initialization stage, both parameter selection and basis enrichment rely exclusively on the HF model.

\subsubsection{High- and Low-Fidelity Snapshot Database Construction}

After the HF greedy procedure is complete, the selected training parameters are fixed. The HF snapshots collected during greedy 
sampling form the HF snapshot matrix
\begin{equation}
\mathbf{S}_{\mathrm{HF}} = \left[ \mathbf{s}_{\mathrm{HF}}(\boldsymbol{\mu}_1), \ldots, \mathbf{s}_{\mathrm{HF}}
	(\boldsymbol{\mu}_{N_s}) \right]
	\label{eq:SHF}
\end{equation}

The LF model is subsequently evaluated at the same parameters in $\mathcal{P}_{\mathrm{HF}}$, producing the corresponding LF snapshot 
matrix
\begin{equation}
\mathbf{S}_{\mathrm{LF}} = \left[ \mathbf{s}_{\mathrm{LF}}(\boldsymbol{\mu}_1), \ldots, \mathbf{s}_{\mathrm{LF}}
	(\boldsymbol{\mu}_{N_s}) \right]
	\label{eq:SLF}
\end{equation}

The resulting parameter-matched HF–LF snapshot database is
\begin{equation*}
\mathbf{S} = \left\{ \left( \mathbf{s}_{\mathrm{HF}}(\boldsymbol{\mu}_i), \mathbf{s}_{\mathrm{LF}}(\boldsymbol{\mu}_i) \right) 
	\right\}_{i=1}^{N_s}
\end{equation*}
These parameter-matched snapshots provide the basis for the subsequent multifidelity correction. No separate greedy sampling is 
performed for the LF model.

\subsection{High--Low-Fidelity Snapshot Alignment}

Before their discrepancy is evaluated, the HF and LF snapshots must be represented on a common spatial grid and in compatible state 
spaces. An LF snapshot $\mathbf{s}_{\mathrm{LF}}(\boldsymbol{\mu}_i)$ is aligned with the HF state space by applying the spatial 
transfer operator $\mathcal{I}_{\mathrm{LF}\rightarrow\mathrm{HF}}$ in~\eqref{eq:LFtoHF}, followed by the state augmentation 
in~\eqref{eq:stildeLF}.

\subsubsection{Low-to-High-Fidelity Interpolation}

At each training parameter $\boldsymbol{\mu}_i$, the Euler snapshot is interpolated from the LF grid onto the HF grid 
using~\eqref{eq:LFtoHF}. In this work, linear interpolation is used for this transfer, although other operators, including conservative
interpolation~\cite{farrell2009conservative}, may also be used. This operation places the flow variables shared by the Euler and RANS 
models on the HF grid. For other model pairings, the same procedure applies to all compatible variables, and interpolation is 
unnecessary when both models use the same grid.

Related work has constructed POD bases from snapshots computed on different adaptively refined meshes, either by expressing them in a 
common finite element space or by evaluating their inner products directly, without interpolation onto a common 
grid~\cite{ullmann2016pod,grassle2018pod,grassle2019pod}. Here, by contrast, interpolation onto the HF grid is required because the 
aligned LF state serves as the reference state for the HF-residual-based correction described in Section~\ref{sec:correction}.

\subsubsection{State Augmentation}
\label{sec:state_augmentation}

For the RANS--Spalart--Allmaras/Euler pairing, interpolation aligns the shared flow variables but does not provide the 
Spalart--Allmaras working variable required by the HF state. The interpolated Euler state is therefore augmented with a zero-valued 
block according to~\eqref{eq:stildeLF}, where the zero block occupies the position of the Spalart--Allmaras working variable in the HF 
state vector. This zero-padding indicates that the LF model contains no corresponding turbulence variable; it does not represent a 
physically zero turbulence field. The resulting discrepancy is represented in the correction ROB constructed in Section~\ref{sec:CB}.

For other model pairings, augmentation depends on which HF state variables have no compatible counterpart in the LF state. For 
example, an Euler LF model paired with a $k$--$\omega$ or $k$--$\varepsilon$ RANS HF model requires zero-valued blocks for both 
turbulence variables. No augmentation is required when both models use the same state variables, even if their grids differ.

When both the HF and LF models include turbulence closures, the corresponding turbulence variables must be treated according to how 
the two closures are defined. If the same turbulence closure is used in both models, the fidelity difference can arise from the 
spatial discretization, with the HF and LF models using different grid resolutions. In this case, the corresponding turbulence 
variables have the same physical definition and can be directly aligned after spatial interpolation. If different turbulence closures 
are used, their turbulence variables may have different physical definitions and should not be directly matched. A variable with no 
physically compatible counterpart in the LF model is instead represented by a zero-valued block in the aligned LF state. State 
alignment must therefore be based on the physical meaning of each variable, not merely its position in the state vector.

Applying the complete state-alignment operator $\mathcal{A}_{\mathrm{LF}\rightarrow\mathrm{HF}}$ yields~\eqref{eq:stildeLF2}.

\subsection{Construction of the High--Low-Fidelity Correction Basis}
\label{sec:CB}

The multifidelity formulation adopted here represents the HF solution as an additive correction to the aligned LF solution. 
Specifically, for a given parameter $\boldsymbol{\mu}$, the HF state is approximated as
\begin{equation}
\mathbf{s}_{\mathrm{HF}}(\boldsymbol{\mu}) \approx \widetilde{\mathbf{s}}_{\mathrm{LF}}(\boldsymbol{\mu}) + 
	\mathbf{V}\mathbf{y}(\boldsymbol{\mu})
\label{eq:shf}
\end{equation}
where $\mathbf{V}$ is a basis for the HF--LF correction space and $\mathbf{y}(\boldsymbol{\mu})$ contains the corresponding correction
coordinates. The construction of $\mathbf{V}$ consists of two steps: forming a discrepancy matrix from parameter-matched snapshots, 
and constructing an orthonormal basis for its column space.

\subsubsection{High--Low-Fidelity Discrepancy Matrix}

Following interpolation~\eqref{eq:LFtoHF} and state augmentation~\eqref{eq:stildeLF}, the paired HF and LF snapshots belong to the 
same algebraic space. For each training parameter $\boldsymbol{\mu}_i$, the HF--LF discrepancy snapshot is defined as
\begin{equation*}
\mathbf{m}_i = \mathbf{s}_{\mathrm{HF}}(\boldsymbol{\mu}_i) - \widetilde{\mathbf{s}}_{\mathrm{LF}}(\boldsymbol{\mu}_i)
\end{equation*}

From~\eqref{eq:SHF} and~\eqref{eq:SLF}, collecting these discrepancy snapshots yields the discrepancy snapshot matrix:
\begin{equation*}
\mathbf{M} = \left[ \mathbf{m}_1,\ldots,\mathbf{m}_{N_s} \right] = \mathbf{S}_{\mathrm{HF}} - \widetilde{\mathbf{S}}_{\mathrm{LF}} 
	\in \mathbb{R}^{N_{\mathrm{HF}}\times N_s}
\end{equation*}
where $N_{\mathrm{HF}}$ is the dimension of the HF state and $N_s$ is the number of parameter-matched training snapshots. Each column 
therefore represents the HF--LF discrepancy at a common parameter value.

The column space of $\boldsymbol{M}$ defines the correction space. Thus, the correction ROB represents the discrepancies between 
aligned LF states and their HF counterparts rather than the HF states themselves.

\subsubsection{Singular Value Decomposition}

An orthonormal correction basis is constructed from the singular value decomposition of the discrepancy matrix
\begin{equation*}
\mathbf{M} = \mathbf{U}\mathbf{\Sigma}\mathbf{W}^{T} 
\end{equation*}
where the columns of $\mathbf{U}$ are the left singular vectors, $\mathbf{W}$ contains the right singular vectors, and 
$\mathbf{\Sigma}$ contains the singular values in nonincreasing order:
\begin{equation*} 
	\sigma_1 \geq \sigma_2 \geq \cdots \geq \sigma_r > 0, \qquad r=\operatorname{rank}(\mathbf{M})
\end{equation*}

No SVD truncation is applied in the present study; the correction basis therefore spans the full numerical range of the discrepancy 
matrix:
\begin{equation*}
\mathbf{V} = \left[ \mathbf{u}_1,\ldots,\mathbf{u}_r \right] \in \mathbb{R}^{N_{\mathrm{HF}}\times r}, \qquad 
	\operatorname{Range}(\mathbf{V}) = \operatorname{Range}(\mathbf{M})
\end{equation*}
Any discrepancy in this space can therefore be represented as $\boldsymbol{V}\boldsymbol{y}$, where $\boldsymbol{y}\in\mathbb{R}^r$ 
contains the correction coordinates.

Because the full discrepancy space is retained, an SVD is not essential: a rank-revealing QR factorization or another suitable 
orthonormalization procedure could construct a basis for the same space. For the snapshot databases considered here, basis construction
represents a relatively small fraction of the overall computational cost compared with the generation of the simulation snapshots. 
Moreover, this cost is amortized over repeated parametric evaluations.

The SVD also permits truncation of the correction ROB if the discrepancy database grows substantially. If a reduced basis of 
dimension $\ell<r$ is desired, the leading left singular vectors define
\begin{equation*}
\mathbf{V}_{\ell} = \left[ \mathbf{u}_1,\ldots,\mathbf{u}_{\ell} \right]
\end{equation*}
The corresponding rank-$\ell$ approximation is
\begin{equation*}
\mathbf{M}_{\ell} = \mathbf{U}_{\ell} \mathbf{\Sigma}_{\ell} \mathbf{W}_{\ell}^{T}
\end{equation*}
with approximation error
\begin{equation*}
\left\| \mathbf{M} - \mathbf{M}_{\ell} \right\|_F^2 = \sum_{j=\ell+1}^{r} \sigma_j^2
\end{equation*}
The singular values can therefore guide the selection of a smaller correction space when storage or reduced-order solution costs 
become significant.

\subsection{Greedy Enrichment of the Multifidelity Reduced-Order Basis}

\subsubsection{Low-Fidelity Prediction}

At the newly selected parameter $\boldsymbol{\mu}^{\star}$, the LF model is evaluated to obtain $\mathbf{s}_{\mathrm{LF}}
(\boldsymbol{\mu}^{\star})$, avoiding a conventional full HF simulation. The resulting LF snapshot is aligned via 
$\mathcal{A}_{\mathrm{LF}\rightarrow\mathrm{HF}}$:
\begin{equation}
\widetilde{\mathbf{s}}_{\mathrm{LF}}(\boldsymbol{\mu}^{\star}) = \mathcal{A}_{\mathrm{LF}\rightarrow\mathrm{HF}} 
	\left( \mathbf{s}_{\mathrm{LF}}(\boldsymbol{\mu}^{\star}) \right)
	\label{eq:ARS}
\end{equation}

The aligned LF solution~\eqref{eq:ARS} serves as the reference state for the HF-residual-based correction. The approximate HF state is
sought in the affine trial space
\begin{equation}
\mathcal{S}(\boldsymbol{\mu}^{\star}) = \widetilde{\mathbf{s}}_{\mathrm{LF}}(\boldsymbol{\mu}^{\star}) + 
	\operatorname{Range}(\mathbf{V})
\label{eq:ats}
\end{equation}

\subsubsection{High-Fidelity-Residual-Based Multifidelity Correction}
\label{sec:correction}

The correction coordinates are determined by minimizing the HF residual over the affine trial space~\eqref{eq:ats}:
\begin{equation*}
\mathbf{y}^{\star} = \underset{\mathbf{y}\in\mathbb{R}^{r}}{\arg\min} \; \frac{1}{2} \left\| \mathbf{R}_{\mathrm{HF}} \left( 
	\widetilde{\mathbf{s}}_{\mathrm{LF}}(\boldsymbol{\mu}^{\star}) + \mathbf{V}\mathbf{y}; \boldsymbol{\mu}^{\star} \right) 
	\right\|_2^2
\label{eq:ystar}
\end{equation*}
where $r$ is the dimension of the correction ROB. The resulting multifidelity (MF) approximation is
\begin{equation}
\mathbf{s}_{\mathrm{MF}}(\boldsymbol{\mu}^{\star}) = \widetilde{\mathbf{s}}_{\mathrm{LF}}(\boldsymbol{\mu}^{\star}) + 
	\mathbf{V}\mathbf{y}^{\star}
\label{eq:smf}
\end{equation}

Although the reference state in~\eqref{eq:ats} is provided by the LF model, the correction is determined by minimizing the residual of
the HF governing equations. Because the LF reference state is evaluated at the query parameter, it accommodates the corresponding 
prescribed far-field state. In the present configuration, the HF and aligned LF solutions share the same prescribed far-field state at 
each paired training parameter $\boldsymbol{\mu}_i$. Their discrepancy therefore vanishes, up to numerical precision, at the shared 
far-field dofs.

Let $\mathbf{B}_{\infty}$ denote the extraction operator for the shared flow variables at these prescribed far-field dofs. Thus, each 
discrepancy snapshot satisfies
\begin{equation*}
\mathbf{B}_{\infty} \left( \mathbf{s}_{\mathrm{HF}}(\boldsymbol{\mu}_i) - \widetilde{\mathbf{s}}_{\mathrm{LF}}(\boldsymbol{\mu}_i) 
	\right) \approx \mathbf{0}, \quad i = 1, \dots, N_s
\end{equation*}
Because the correction basis $\mathbf{V}$ is constructed directly from the span of these discrepancy snapshots, it inherits this 
boundary property, yielding
\begin{equation}
\mathbf{B}_{\infty}\mathbf{V} \approx \mathbf{0}
\label{eq:prop1}
\end{equation}

Equations~\eqref{eq:smf} and~\eqref{eq:prop1} then imply that, provided interpolation preserves the prescribed far-field values,
\begin{equation*}
\mathbf{B}_{\infty} \mathbf{s}_{\mathrm{MF}} (\boldsymbol{\mu}^{\star}) \approx \mathbf{B}_{\infty} 
	\widetilde{\mathbf{s}}_{\mathrm{LF}} (\boldsymbol{\mu}^{\star}) = \mathbf{g}_{\infty}(\boldsymbol{\mu}^{\star})
\end{equation*}
where $\mathbf{g}_{\infty}$ denotes the shared prescribed far-field state. Farfield preservation is exact if the correction basis 
vanishes on these dofs and otherwise holds approximately.

This argument does not guarantee satisfaction of the HF wall boundary conditions, which differ between the Euler and RANS models. The 
need for further HF RANS smoothing is assessed in Section~\ref{sec:RBS}.

\subsection{Residual-Based RANS Smoothing}
\label{sec:RBS}

\subsubsection{RANS Residual Cutoff}

The corrected state $\mathbf{s}_{\mathrm{MF}}(\boldsymbol{\mu}^{\star})$ is assessed before being accepted into the PROM snapshot 
database. Its HF residual norm is defined as
\begin{equation*}
\rho_{\mathrm{MF}}(\boldsymbol{\mu}^{\star}) = \left\| \mathbf{R}_{\mathrm{HF}} \left(\mathbf{s}_{\mathrm{MF}}
	(\boldsymbol{\mu}^{\star}); \boldsymbol{\mu}^{\star} \right) \right\|_2
\end{equation*}

The candidate residual is compared with the HF residual of the current PROM prediction at the same parameter. Denoting this prediction
by $\mathbf{s}_{\mathrm{PROM}}(\boldsymbol{\mu}^{\star})$, its residual is
\begin{equation*}
\rho_{\mathrm{PROM}}(\boldsymbol{\mu}^{\star}) = \left\| \mathbf{R}_{\mathrm{HF}} \left(\mathbf{s}_{\mathrm{PROM}} 
	(\boldsymbol{\mu}^{\star}); \boldsymbol{\mu}^{\star} \right) \right\|_2
\end{equation*}
Both residual norms use the same definition and scaling.

If
\begin{equation*}
\rho_{\mathrm{MF}}(\boldsymbol{\mu}^{\star}) < \rho_{\mathrm{PROM}}(\boldsymbol{\mu}^{\star})
\end{equation*}
the corrected multifidelity solution has a lower HF residual than the current PROM prediction and is accepted directly into the PROM 
database.

If, instead,
\begin{equation*}
\rho_{\mathrm{MF}}(\boldsymbol{\mu}^{\star}) \geq \rho_{\mathrm{PROM}}(\boldsymbol{\mu}^{\star})
\end{equation*}
RANS smoothing is applied to reduce its HF residual before reconsidering it for enrichment.

This criterion assesses each candidate locally: a lower HF residual supports acceptance but does not guarantee improved PROM accuracy 
throughout the parameter domain.

\subsubsection{RANS Smoothing Criterion}

The corrected state in~\eqref{eq:smf} initializes the HF RANS solver:
\begin{equation*}
\mathbf{s}^{(0)} = \mathbf{s}_{\mathrm{MF}}(\boldsymbol{\mu}^{\star}) 
\end{equation*}
The HF RANS solver is then advanced according to
\begin{equation*}
\mathbf{s}^{(j+1)} = \mathcal{G}_{\mathrm{HF}} \left( \mathbf{s}^{(j)}; \boldsymbol{\mu}^{\star} \right)
\end{equation*}
where $\mathcal{G}_{\mathrm{HF}}$ denotes the HF RANS solution operator that advances the state by one iteration at 
$\boldsymbol{\mu}^{\star}$.

Smoothing continues until the HF residual norm reaches the prescribed target $\rho_{\mathrm{cut}}$. Specifically, starting from 
$\mathbf{s}^{(0)}=\mathbf{s}_{\mathrm{MF}}(\boldsymbol{\mu}^{\star})$, RANS iterations continue until
\begin{equation*}
\left\| \mathbf{R}_{\mathrm{HF}} \left( \mathbf{s}^{(j)}; \boldsymbol{\mu}^{\star} \right) \right\|_2 \leq \rho_{\mathrm{cut}}
\end{equation*}
The target may be defined as a fraction of the residual norm of the current PROM prediction:
\begin{equation*}
\rho_{\mathrm{cut}} = \beta\, \rho_{\mathrm{PROM}}(\boldsymbol{\mu}^{\star}), \qquad 0 < \beta < 1
\end{equation*}
where the prescribed parameter $\beta$ determines the required reduction relative to the current PROM residual.

The procedure thus combines residual-based parameter selection, LF-based correction in an HF-residual-informed affine space, and 
selective RANS smoothing. Each accepted snapshot enriches the final PROM database and updates the correction ROB, reducing reliance 
on conventional fully converged HF simulations during training.

\section{Numerical Results}
\label{sec:NR}

\subsection{Steady-State Transonic Flow Past a NACA0012 Airfoil}

The proposed methodology is evaluated for steady-state transonic flow past a NACA0012 airfoil. The HF model solves the RANS 
equations with the Spalart--Allmaras turbulence model, whereas the LF model solves the Euler equations. The two models are discretized
on separate computational grids, such that the fidelity difference reflects both physical modeling and spatial discretization. In all 
cases, the PROMs are constructed using LSPG projection without hyperreduction.

\subsubsection{Parameter Domain}

The parameter vector comprises the freestream Mach number and angle of attack:
\begin{equation*}
\boldsymbol{\mu}=(M_\infty,\alpha)
\end{equation*}
The parameter domain is defined as
\begin{equation*}
\mathcal{D}= [0.4,0.65]\times[0^\circ,5^\circ]
\end{equation*}
The upper bound on the Mach number is restricted to $0.65$ to avoid the onset of transonic shock-induced buffet, a phenomenon in which 
shock motion and boundary-layer separation can produce self-sustained, unsteady aerodynamic loads. The airfoil geometry remains fixed 
throughout the study, and the HF and LF models are evaluated at matching parameter values. 

\subsubsection{Training and Test Parameter Sets}

A uniform tensor-product grid with $22$ points in each parameter direction defines the greedy candidate set:
\begin{equation*}
\mathcal{D}_{\mathrm{cand}} = \left\{ \left(0.4 + \tfrac{0.25}{21}\,i,\; \tfrac{5}{21}\,j^\circ\right) : i,j = 0,\ldots,21 \right\}
\end{equation*}
This set contains $484$ candidate parameter points, from which parameters are selected by the residual-based greedy procedure of 
Section~\ref{sec:greedy}; an HF snapshot is not generated at every candidate point. During multifidelity enrichment, the PROM used to 
evaluate the residual indicator is constructed from the combined database of HF and accepted corrected-LF snapshots.

Prediction accuracy is assessed at $50$ test parameter points in $\mathcal{D}$, sampled independently of the candidate grid used for 
greedy parameter selection. The sampling of these points is described in Section~\ref{sec:evaluation}.

\subsubsection{High-Fidelity RANS Model}

\begin{figure}[htbp]
\centering
\begin{minipage}[t]{0.43\textwidth}
    \centering
    \includegraphics[width=\linewidth]{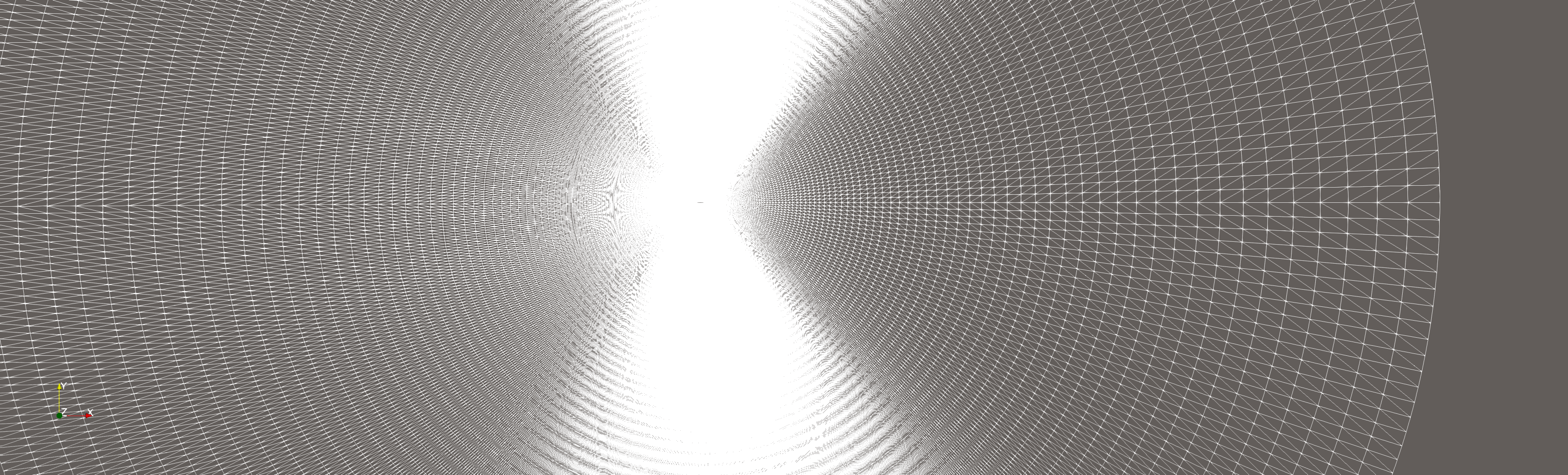}
    \par\smallskip
    (a) Full discretized computational domain
\end{minipage}
\hfill
\begin{minipage}[t]{0.53\textwidth}
    \centering
    \includegraphics[width=\linewidth]{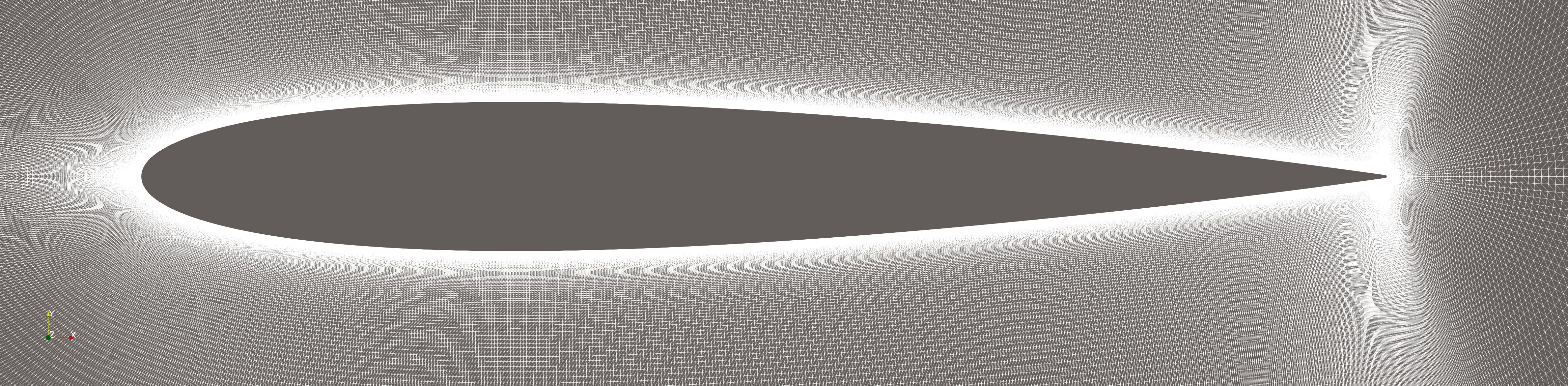}
    \par\smallskip
    (b) Near-airfoil mesh
\end{minipage}
\caption{Computational domain and mesh of the HF model: (a) full domain with a circular far-field boundary located approximately 
	$80$~chord lengths from the airfoil; (b) close-up of the airfoil surface showing the near-wall mesh refinement.}
\label{fig:rans_domain}
\end{figure}

The computational domain of the HF model is bounded by the airfoil surface and a circular far-field boundary located approximately 
$80$~chord lengths from the airfoil. Figure~\ref{fig:rans_domain} shows the full computational domain and a close-up of the 
near-airfoil mesh.

The HF model solves the steady RANS equations with the Spalart--Allmaras turbulence model. Its computational grid contains 
$812\,098$ nodes and includes near-wall refinement for wall-resolved calculations, with a wall resolution of $y^+\approx0.75$.

The HF model generates the initial directly computed training snapshots and the reference solutions used for accuracy assessment. Its 
discrete residual also provides the greedy sampling indicator and the objective for multifidelity correction. The same HF solver is 
subsequently used to smooth corrected LF predictions before they are accepted into the final PROM training database.

\subsubsection{Low-Fidelity Euler Model}

The LF model solves the steady Euler equations on a separate coarse grid containing approximately $54\,000$ nodes. Thus, the HF and 
LF models differ in both their governing physical models and their spatial discretizations.

LF solutions are initially computed at the parameters selected by the HF greedy procedure to form parameter-matched HF--LF snapshot 
pairs. During subsequent multifidelity enrichment, an LF solution is computed at each newly selected parameter. The solution is then 
linearly interpolated onto the HF grid and augmented with a zero-valued Spalart--Allmaras variable before HF-residual correction and 
RANS smoothing.

\subsubsection{High-Fidelity Snapshot Database}

The initial HF sampling is seeded with the four corners of the two-dimensional parameter domain:
\begin{equation}
\mathcal{P}_{\mathrm{HF}}^{(4)}= \left\{ (0.4,0^\circ),\, (0.4,5^\circ),\, (0.65,0^\circ),\, (0.65,5^\circ) \right\}
\label{eq:P4}
\end{equation}
Fully converged HF solutions at these parameter values form the smallest initial HF snapshot database.

Larger initial databases are obtained by continuing the standard HF residual-based greedy procedure.

\subsection{High-Fidelity PROM and Evaluation Metrics}
\label{sec:evaluation}

\begin{figure}[htbp]
\centering
\includegraphics[width=0.5\textwidth]{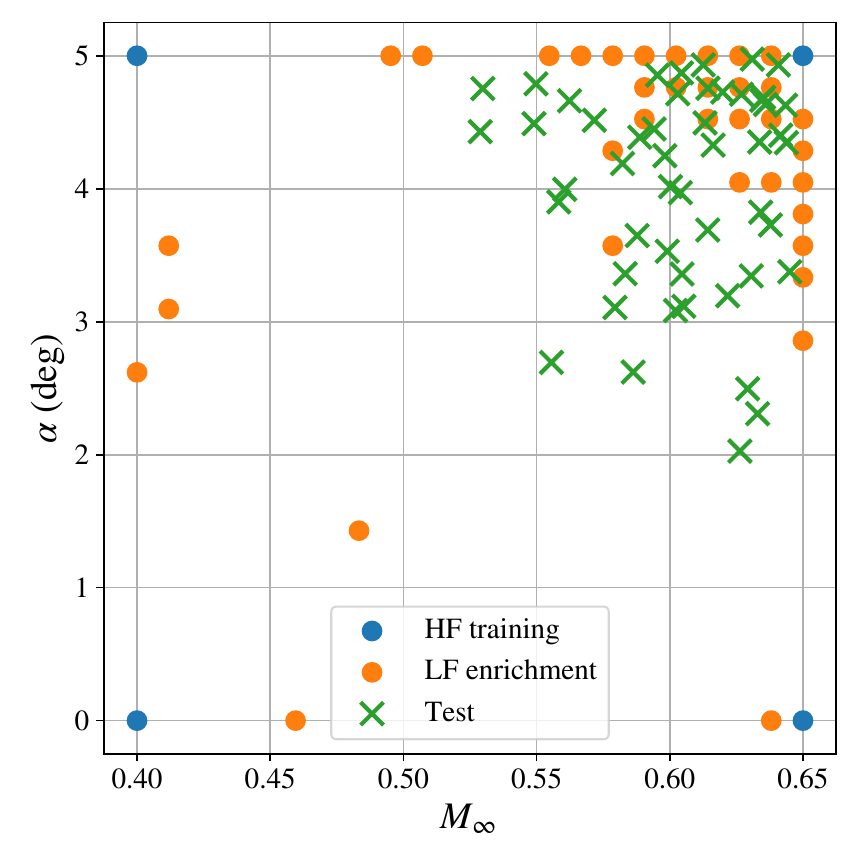}
\caption{Parameter domain showing the HF training points and LF enrichment points of MF-PROM4, and the off-training test points.}
\label{fig:parameter_space_mf}
\end{figure}

The HF-PROM -- that is, the LSPG PROM constructed from up to $40$ HF snapshots selected by the HF greedy procedure -- and the MF-PROMs
are evaluated against fully converged HF-HDM solutions at $N_{\mathrm{test}} = 50$ off-training parameter points. The same test set is
used for all PROMs, including the MF-PROMs initialized with different numbers of HF snapshots, and for all ROB dimensions, denoted by 
$n$. Figure~\ref{fig:parameter_space_mf} shows the parameter domain, the locations of the HF training points and LF enrichment points 
of MF-PROM4, and the test points.

The test points are sampled with a bias toward higher freestream Mach numbers and angles of attack, all lying within $[0.53, 0.65] 
\times [2^\circ, 5^\circ]$, which emphasizes the shock-containing region of $\mathcal{D}$ and provides a more demanding assessment of 
prediction accuracy for flows whose shock positions vary with the parameters.

Accuracy is assessed using percentage relative errors in the lift ($L$), drag ($D$), and lift-to-drag ratio ($L/D$), together with a 
surface-integrated relative $L^2$ error in the streamwise skin-friction coefficient. Unless otherwise stated, these errors are 
averaged over the $N_{\mathrm{test}}$ test points. The reported mean errors therefore characterize performance over the biased test 
set rather than uniform averages over $\mathcal{D}$.

\newpage
\subsection{Prediction Accuracy}

This section compares the prediction accuracy of the HF-PROM and the MF-PROMs. The MF-PROM constructed from the $k$ directly computed
HF snapshots at the parameter points in $\mathcal{P}_{\mathrm{HF}}^{(k)}$, together with the LF-derived snapshots generated during 
enrichment, is denoted by MF-PROM$k$. The figures accordingly label the MF-PROMs initialized with $k = 4, 6, \dots, 12$ HF snapshots 
as MF-PROM4, MF-PROM6, \dots, MF-PROM12. Unless otherwise stated, the MF-PROM considered is MF-PROM4, whose four HF snapshots are 
computed at the parameter points in $\mathcal{P}_{\mathrm{HF}}^{(4)}$~\eqref{eq:P4}.

\subsubsection{Aerodynamic Performance}

\begin{figure}[h!]
\centering
\includegraphics[width=0.985\textwidth]{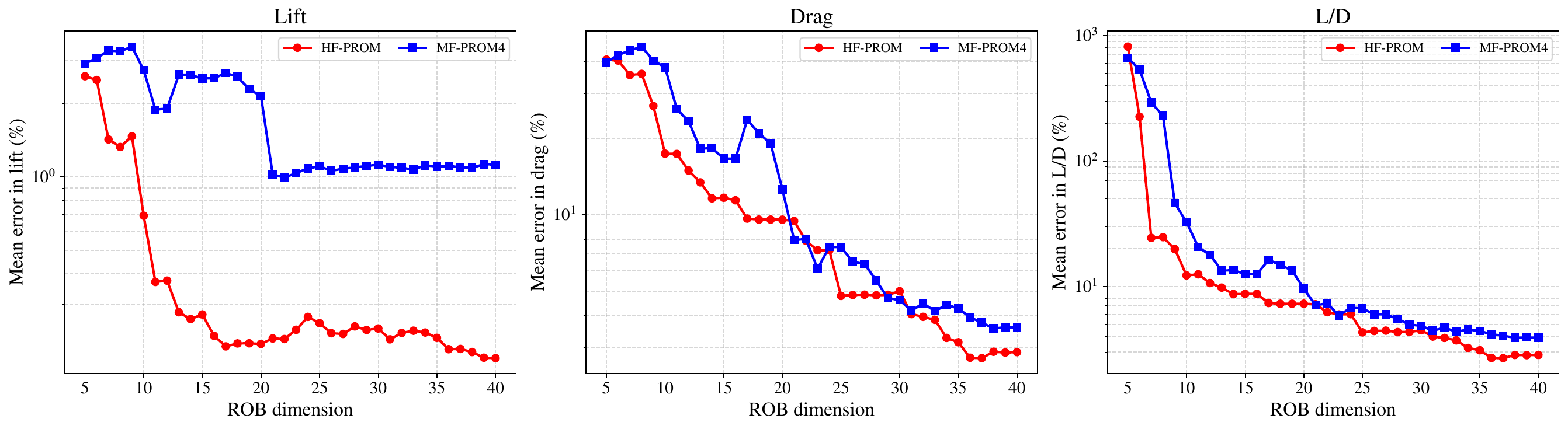}
\caption{Mean percentage errors in lift, drag, and lift-to-drag ratio predicted by the HF-PROM and the MF-PROM4, relative to the 
	HF-HDM reference, as functions of the ROB dimension.}
\label{fig:aerodynamic_accuracy}
\end{figure}

Figure~\ref{fig:aerodynamic_accuracy} shows the mean percentage errors in lift, drag, and lift-to-drag ratio predicted by the
HF-PROM and MF-PROM4 as functions of the ROB dimension $N$. For both PROMs, the errors generally decrease as the ROB dimension
increases, although the convergence is non-monotonic. Beyond $N = 20$, the MF-PROM4 lift error fluctuates near $1.1\%$,
reaching $1.12\%$ at $N = 40$, whereas the HF-PROM lift error decreases overall to $0.18\%$.

At the largest ROB dimension considered, $N = 40$, the differences between the two PROMs are smaller for drag and lift-to-drag ratio
than for lift. The mean drag errors of the HF-PROM and MF-PROM4 are $2.88\%$ and $3.59\%$, respectively, while their mean
lift-to-drag ratio errors are $2.84\%$ and $3.90\%$.  Thus, at this dimension, MF-PROM4 achieves mean errors below
$4\%$ in all three aerodynamic quantities, but is less accurate than the HF-PROM in each quantity. These mean errors characterize
the aggregate prediction accuracy over the evaluated test set; they do not establish a maximum-error bound or compliance with
an application-specific accuracy requirement.

\subsubsection{Skin-Friction Coefficient}

\begin{figure}[h!]
\centering
\includegraphics[width=0.8\textwidth]{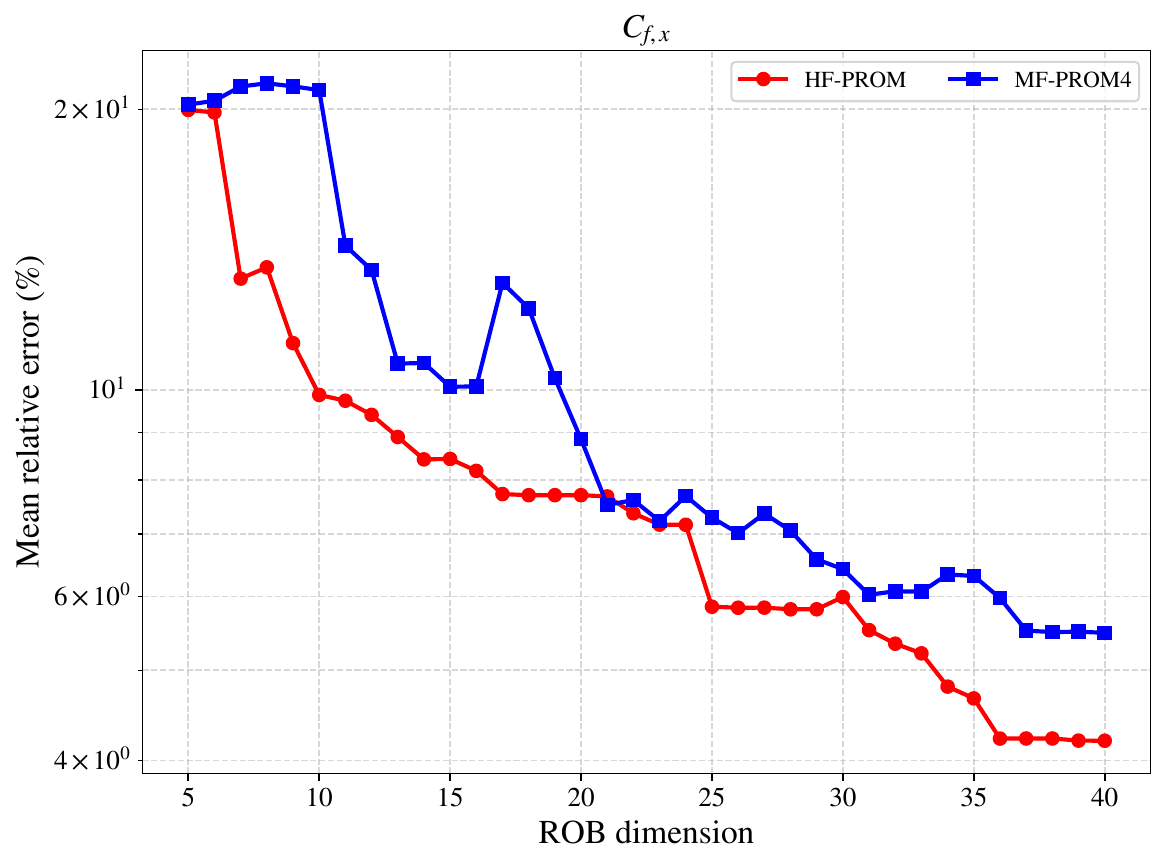}
\caption{Mean surface-integrated relative $L^2$ error in the streamwise skin-friction coefficient, expressed as a percentage, versus 
	ROB dimension for the HF-PROM and MF-PROM4.}
\label{fig:cf_error_convergence}
\end{figure}

Figure~\ref{fig:cf_error_convergence} compares the mean surface-integrated relative $L^2$ error in the streamwise skin-friction 
coefficient, $C_{f,x}$, as a function of the ROB dimension. The results quantify how accurately the HF-PROM and MF-PROM4 reproduce the 
$C_{f,x}$ distributions predicted by the HF-HDM. For both PROMs, the error decreases as the ROB dimension increases, from approximately
$20\%$ at $n = 5$ to approximately $4\%$ for the HF-PROM and $5\%$ for the MF-PROM4 at $n = 40$.

\begin{figure}[h!]
\centering
\includegraphics[width=0.8\textwidth]{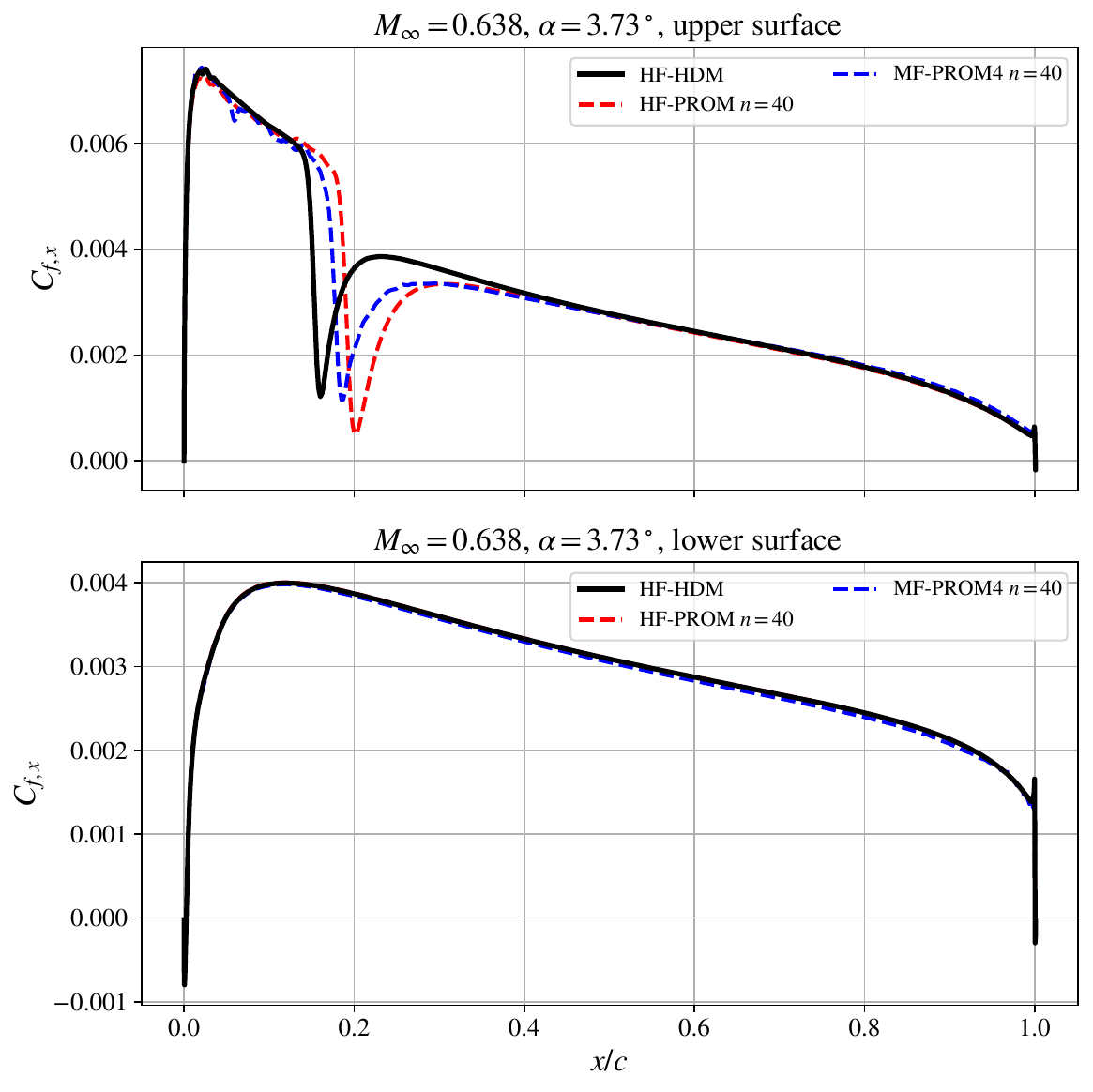}
\caption{Upper- and lower-surface streamwise skin-friction coefficient distributions at a selected test parameter point, comparing the
HF-PROM and MF-PROM4 predictions, both with a ROB of dimension 40, to the HF-HDM reference. The chordwise coordinate $x$ is normalized
by the airfoil chord $c$.}
\label{fig:cf_example}
\end{figure}

Figure~\ref{fig:cf_example} complements these aggregate error measures with a comparison of the upper- and lower-surface $C_{f,x}$ 
distributions at a selected test parameter point. For a ROB of dimension $n=40$, both PROMs capture the overall shape of the sharp 
skin-friction feature associated with the shock but predict it at a location displaced from that of the HF-HDM reference. 
This behavior is consistent with the known difficulty of representing parameter-dependent shock positions in a low-dimensional, fixed
affine trial space~\cite{reiss2018shifted}: a linear combination of spatially fixed modes does not explicitly transport such features,
and accurately representing their changing locations can require a comparatively large basis. 

At this parameter point and ROB dimension, the $C_{f,x}$ distribution predicted by the MF-PROM4 is closer to the HF-HDM reference than 
that predicted by the HF-PROM. This observation is specific to the selected case and should not be interpreted as evidence of 
systematic superiority; the averaged errors in Fig.~\ref{fig:cf_error_convergence} provide the broader assessment of relative accuracy.

\newpage

\subsection{Effect of High-Fidelity Snapshot Count}

\begin{figure}[h!]
\centering
\includegraphics[width=0.995\textwidth]{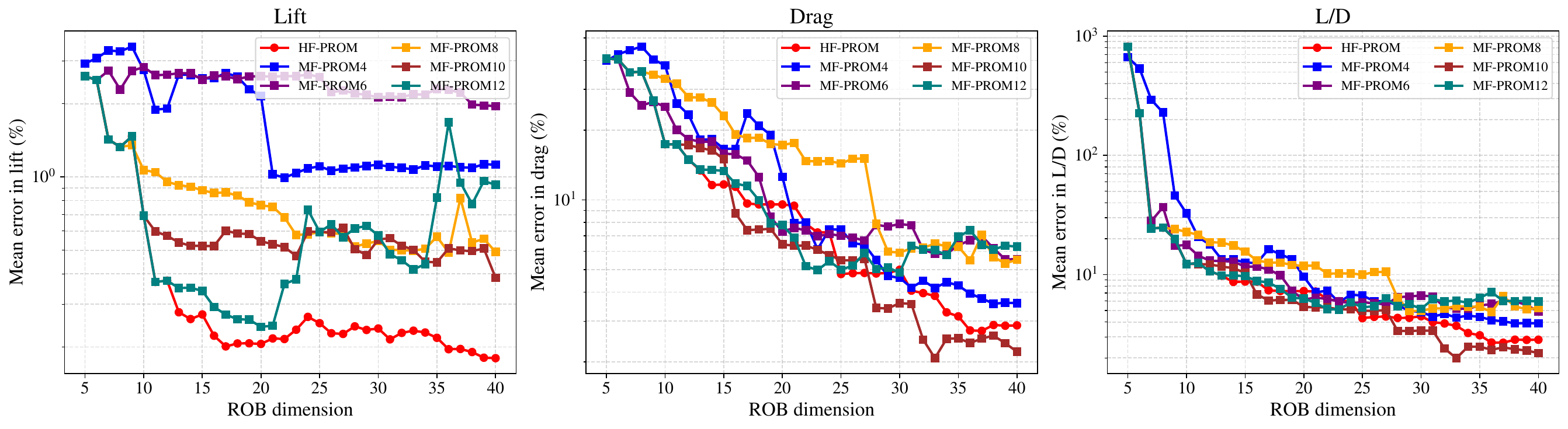}
\caption{Mean percentage errors in lift, drag, and lift-to-drag ratio relative to the HF-HDM reference, versus ROB dimension, for the 
HF-PROM and MF-PROMs initialized with different numbers of HF snapshots.}
\label{fig:L_D_LovD_error_convergence_diff_basis}
\end{figure}

\begin{figure}[h!]
\centering
\includegraphics[width=0.8\textwidth]{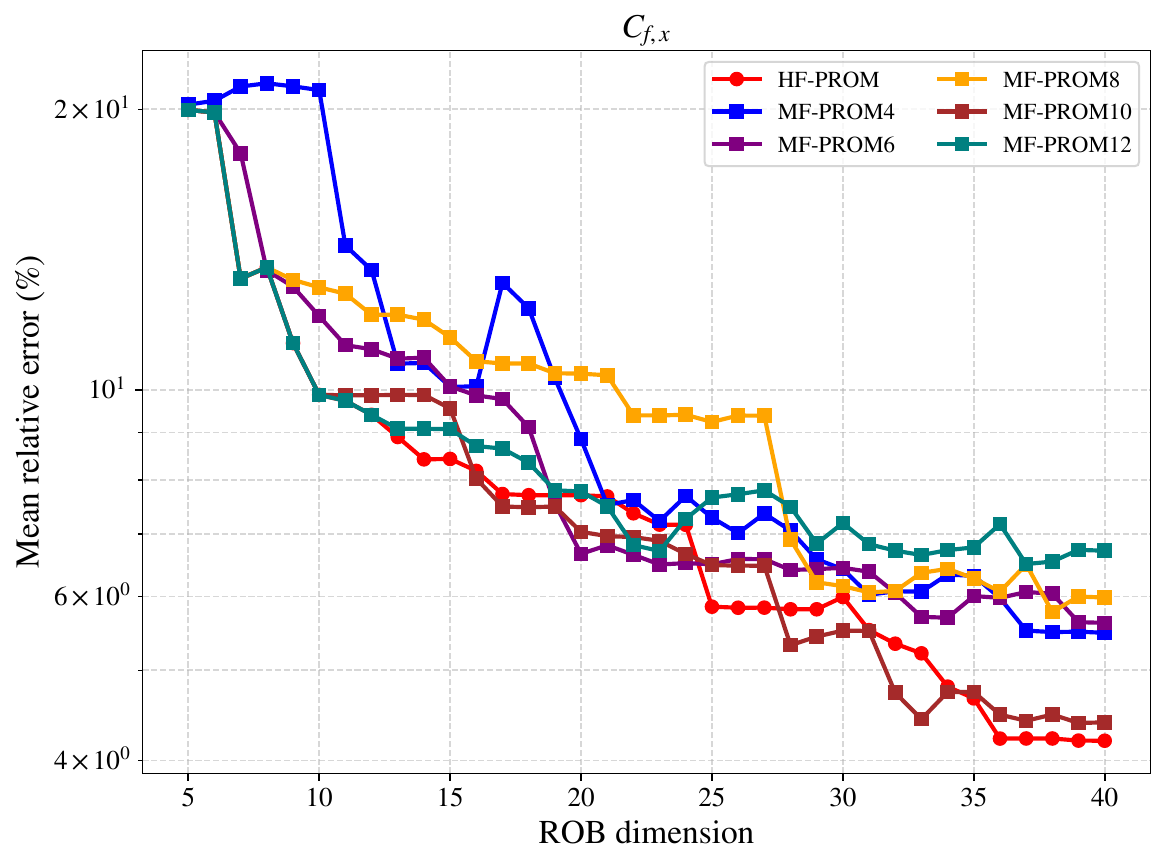}
\caption{Mean surface-integrated relative $L^2$ error in the streamwise skin-friction coefficient, expressed as a percentage, versus 
ROB dimension for the HF-PROM and MF-PROMs initialized with different numbers of HF snapshots.}
\label{fig:cf_error_convergence_diff_basis}
\end{figure}

Next, the number $n_{\mathrm{HF}}$ of directly computed HF snapshots used to initialize the multifidelity construction, that is, $k$ 
in MF-PROM$k$, is varied from $4$ to $12$ in increments of two. The initial databases are nested: each consists of the first 
$n_{\mathrm{HF}}$ snapshots of the greedy sampling sequence used to train the HF-PROM. 
Figures~\ref{fig:L_D_LovD_error_convergence_diff_basis} and~\ref{fig:cf_error_convergence_diff_basis} compare the resulting 
aerodynamic and skin-friction prediction errors as functions of the ROB dimension.

At the largest ROB dimension considered, $n = 40$, the mean errors in the lift, drag, and lift-to-drag ratio predicted by the MF-PROMs
range from $0.39\%$ to $1.96\%$, $2.21\%$ to $6.29\%$, and $2.19\%$ to $5.95\%$, respectively, depending on $n_{\mathrm{HF}}$
(Fig.~\ref{fig:L_D_LovD_error_convergence_diff_basis}). For the skin-friction coefficient, the mean errors of the MF-PROMs range from 
approximately $4\%$ to $7\%$, compared with approximately $4\%$ for the HF-PROM (Fig.~\ref{fig:cf_error_convergence_diff_basis}).

At $n = 40$, MF-PROM10 yields the lowest errors among the tested
MF-PROMs in all three aerodynamic quantities: $0.39\%$ for lift, $2.21\%$ for drag, and 
$2.19\%$ for the lift-to-drag ratio. For the MF-PROM12, these errors increase to $0.93\%$, $6.29\%$, and $5.95\%$, respectively. For 
comparison, the corresponding errors of the HF-PROM predictions are $0.18\%$, $2.88\%$, and $2.84\%$. At this dimension, the MF-PROM10
is therefore more accurate than the HF-PROM in drag and lift-to-drag ratio but less accurate in lift. This observation is specific to 
the selected ROB dimension and should not be interpreted as evidence of systematic superiority since, as discussed below, the relative
accuracy of the PROMs varies with the ROB dimension.

Indeed, the ranking of the MF-PROMs depends on both the ROB dimension and the quantity of interest. For example, at $n = 20$, the 
MF-PROM12 yields the lowest mean lift error among the MF-PROMs ($0.24\%$), whereas the MF-PROM10 yields the lowest mean drag and 
lift-to-drag ratio errors ($6.43\%$ and $5.36\%$, respectively). Consequently, no single value of $n_{\mathrm{HF}}$ is uniformly most 
accurate across the tested ROB dimensions and quantities of interest, and the prediction accuracy of the MF-PROM at a fixed ROB 
dimension does not, in general, improve monotonically with $n_{\mathrm{HF}}$.

This non-monotonic behavior is consistent with the configuration-dependent nature of the adaptive enrichment procedure. At each 
enrichment step, the next parameter is selected by maximizing the HF residual of the current PROM, and therefore depends on the 
snapshots already in the training database. Each LF-derived snapshot is a corrected approximation of the corresponding HF solution, 
with the correction determined by a correction ROB that is initially constructed from the $n_{\mathrm{HF}}$ available discrepancy 
snapshots and subsequently updated with each accepted enrichment. Changing $n_{\mathrm{HF}}$ thus changes the correction space, which 
can alter both the corrected snapshots and the subsequent enrichment sequence. Although the initial HF databases are nested, the final
training databases and the ROBs they span are therefore not guaranteed to be nested, and a larger HF initialization need not produce 
a more accurate MF-PROM. Quantifying how much of the observed non-monotonicity is attributable to the differing enrichment 
trajectories would require comparing the selected parameter sequences and correction errors across configurations.

\subsection{Computational Cost}

All computations were performed on a Linux cluster using $10$ compute nodes with $24$ cores each, for a total of $240$ cores.

The computational cost is evaluated for MF-PROMs initialized with $n_{\mathrm{HF}} \in \{4,6,8,10,12\}$ directly computed HF snapshots.
Each configuration is enriched with $40 - n_{\mathrm{HF}}$ corrected and, when necessary, smoothed LF snapshots so that every final 
training database contains $40$ snapshots in the HF solution space. The corresponding numbers of LF-derived snapshots are $36$, $34$, 
$32$, $30$, and $28$.

For each MF-PROM, the enrichment cost comprises three components: LF-HDM evaluation, multifidelity correction, and RANS smoothing. 
The average wall-clock time of each component is its accumulated time divided by the number of enrichment evaluations,
\begin{equation*}
N_{\mathrm{enrich}} = 40-n_{\mathrm{HF}},
\end{equation*}
and thus represents the cost of generating one additional LF-derived training snapshot. The average enrichment cost is
\begin{equation*}
\overline{T}_{\mathrm{enrich}} = \overline{T}_{\mathrm{LF,HDM}} + \overline{T}_{\mathrm{corr}} + \overline{T}_{\mathrm{smooth}}
\end{equation*}
The component averages are reported in Table~\ref{tab:mf_component_costs}.

\begin{table}[h!]
\centering
\caption{Average wall-clock time per multifidelity enrichment evaluation for different initial HF snapshot counts. Smoothing times are
averaged over all enrichment evaluations, including those for which no smoothing was required. Totals may differ from the sums of the 
displayed components because of rounding.}
\label{tab:mf_component_costs}
\begin{tabular}{rrrrrr}
\toprule
Initial HF & LF-derived & LF-HDM & Correction & Smoothing & Total \\
snapshots  & snapshots  & [s]    & [s]        & [s]       & [s]   \\
\midrule
4  & 36 & 8.9 & 7.3 & 28.0 & 44.2 \\
6  & 34 & 5.9 & 4.4 & 22.5 & 32.9 \\
8  & 32 & 5.4 & 3.1 & 18.4 & 27.0 \\
10 & 30 & 7.7 & 4.7 & 25.9 & 38.4 \\
12 & 28 & 6.4 & 3.8 & 21.4 & 31.6 \\
\bottomrule
\end{tabular}
\end{table}

The average LF-HDM evaluation time ranges from $5.4$ to $8.9~\mathrm{s}$ and does not decrease monotonically with $n_{\mathrm{HF}}$. 
This behavior is expected because the configurations select different enrichment parameters and therefore need not incur identical LF 
solution costs. The average multifidelity correction time ranges from $3.1$ to $7.3~\mathrm{s}$, with the largest value obtained for
$n_{\mathrm{HF}}=4$. The average RANS smoothing time ranges from $18.4$ to $28.0~\mathrm{s}$ and is the largest of the three enrichment
components in every configuration.

The corrected LF prediction initializes the HF RANS solver, which is advanced until the prescribed residual criterion is satisfied. 
Unlike a conventionally initialized HF simulation advanced to full convergence, this step refines an already corrected prediction and 
may accept snapshots at different residual levels. Consequently, the cost comparison must be interpreted together with the 
prediction-accuracy results.

Including the directly computed HF initialization, the total snapshot-generation cost is
\begin{equation*}
T_{\mathrm{MF}}(n_{\mathrm{HF}}) = T_{\mathrm{HF,init}}(n_{\mathrm{HF}}) + T_{\mathrm{enrich,total}}(n_{\mathrm{HF}})
\end{equation*}
where
\begin{equation*}
T_{\mathrm{enrich,total}} = N_{\mathrm{enrich}}\,\overline{T}_{\mathrm{enrich}}  
\end{equation*}
The initialization cost increases from $4801~\mathrm{s}$ for $n_{\mathrm{HF}}=4$ to $14\,241~\mathrm{s}$ for $n_{\mathrm{HF}}=12$, 
whereas the aggregate enrichment cost ranges from $863$ to $1590~\mathrm{s}$. Table~\ref{tab:mf_total_costs} summarizes the 
initialization, enrichment, and total costs. Although larger initial HF databases require fewer LF-derived snapshots, the additional 
cost of directly computing HF snapshots outweighs the savings in enrichment, causing the total snapshot-generation cost to increase
with $n_{\mathrm{HF}}$.

\subsection{Computational Speedup}

The HF-only construction requires $49\,858~\mathrm{s}$ to generate $40$ directly computed HF snapshots, or approximately 
$1\,246~\mathrm{s}$ per snapshot. The snapshot-generation speedup is defined as
\begin{equation*}
S(n_{\mathrm{HF}}) = \frac{T_{\mathrm{HF\text{-}only}}}{T_{\mathrm{MF}}(n_{\mathrm{HF}})}
\end{equation*}
For $n_{\mathrm{HF}} = 4$, the total multifidelity cost is $6\,392~\mathrm{s}$, yielding $S(4) \approx 7.80$. For 
$n_{\mathrm{HF}} = 6$, $8$, $10$, and $12$, the speedups are $6.10$, $4.91$, $3.88$, and $3.30$, respectively 
(Table~\ref{tab:mf_total_costs}).

These speedups compare snapshot-generation procedures under their respective convergence criteria, not at identical final residual 
levels. They also refer to the offline snapshot-generation cost only and exclude subsequent online PROM evaluations.

\begin{table}[htbp]
\centering
\caption{Aggregate snapshot-generation costs and speedups relative to the HF-only baseline of $49\,858~\mathrm{s}$. Each training 
	database contains $40$ snapshots. Totals may differ from the sums of the displayed components because of rounding.}
\label{tab:mf_total_costs}
\begin{tabular}{rrrrr}
\toprule
Initial HF & HF initialization & MF enrichment & MF total & Speedup \\
snapshots  & [s]               & [s]           & [s]      &         \\
\midrule
4  & 4\,801   & 1\,590 & 6\,392   & 7.80 \\
6  & 7\,058   & 1\,119 & 8\,176   & 6.10 \\
8  & 9\,288   & 863  & 10\,151 & 4.91 \\
10 & 11\,695 & 1\,152 & 12\,847 & 3.88 \\
12 & 14\,241 & 884  & 15\,124 & 3.30 \\
\bottomrule
\end{tabular}
\end{table}

\subsection{Accuracy--Cost Tradeoff}
\label{sec:accuracy_cost}

Increasing the initial HF snapshot count introduces a tradeoff between prediction accuracy and training cost. A larger initial HF 
database supplies additional high-fidelity information for constructing both the final PROM ROB and the correction ROB, but at a 
greater initial computational cost.

Among the configurations examined, $n_{\mathrm{HF}} = 4$ yields the lowest snapshot-generation cost and the largest speedup, whereas 
$n_{\mathrm{HF}} = 10$ yields the best overall aerodynamic prediction accuracy at a ROB dimension of $n = 40$. The accuracy gains are 
not uniform across the quantities of interest: lift shows the clearest improvement, drag and lift-to-drag ratio exhibit less consistent
trends, and increasing $n_{\mathrm{HF}}$ from $10$ to $12$ degrades the accuracy in all three aerodynamic quantities at $n = 40$, 
despite the higher initialization cost. This ranking, however, varies with the ROB dimension: at $n = 20$, for example, 
$n_{\mathrm{HF}} = 12$ yields the lowest lift error among the MF-PROMs. Comparisons at a single ROB dimension should therefore be 
interpreted with caution; the results indicate only that a larger HF initialization does not reliably improve accuracy.

The preferred initialization therefore depends on the required accuracy, the quantities of interest, and the ROB dimension. A small 
initial HF database is sufficient when its prediction errors meet the application requirements, whereas a larger one is justified only
when it demonstrably improves accuracy enough to warrant the additional training cost. In the present study, $n_{\mathrm{HF}} = 4$ is 
the cost-oriented choice, and $n_{\mathrm{HF}} = 10$ is the most accurate configuration at $n = 40$; neither is universally optimal 
in the absence of a prescribed accuracy requirement.


\section{Conclusions}
\label{sec:C}

This work presented a multifidelity methodology for reducing the offline cost of constructing the reduced-order basis (ROB) of a
projection-based reduced-order model (PROM) for computational fluid dynamics (CFD). The methodology accounts for fidelity discrepancies
arising from the governing physical model, the spatial discretization, or both, and aligns low-fidelity (LF) and high-fidelity (HF)
solutions through spatial interpolation and physically consistent state augmentation. Discrepancies between parameter-matched HF and
aligned LF snapshots define a correction ROB, within which LF predictions at new parameter points are corrected by minimizing
the HF residual over an affine trial space. When the corrected state does not sufficiently reduce the HF residual, a short
Reynolds-averaged Navier--Stokes (RANS) smoothing stage is applied. Each accepted snapshot enriches both the PROM training database
and the correction ROB. Because the methodology generates training data rather than modifying the projection itself, the resulting
ROB can be used with established projection-based model order reduction formulations, including least-squares Petrov--Galerkin
(LSPG) projection.

For steady compressible flow over a NACA0012 airfoil, with a fine-grid RANS model as the HF model and a coarse-grid Euler model
as the LF model, the methodology constructed a $40$-snapshot training database from four directly computed HF snapshots and
$36$ corrected and smoothed LF-derived snapshots. Averaged over the LF-derived snapshots, generation required approximately
$44$~s per snapshot, including an average RANS smoothing contribution of approximately $28$~s; smoothing was applied only
when required. A fully converged HF snapshot required approximately $1\,246$~s on average. The total snapshot-generation
cost was reduced by a factor of $7.8$ relative to HF-only construction.

The resulting multifidelity PROM, denoted MF-PROM4 to indicate its four directly computed HF snapshots, and a PROM constructed
entirely from HF snapshots (HF-PROM) both used LSPG projection with a ROB of dimension $40$. MF-PROM4 predicted lift, drag, and
lift-to-drag ratio with mean errors of approximately $1.1\%$, $3.6\%$, and $3.9\%$, respectively, compared with $0.18\%$,
$2.9\%$, and $2.8\%$ for the HF-PROM. The corresponding mean relative errors in the streamwise skin-friction coefficient were
approximately $5\%$ and $4\%$, respectively. These errors were obtained over a test set deliberately biased toward the
shock-containing region of the parameter domain. Thus, MF-PROM4 reduced the snapshot-generation cost while incurring higher mean
errors than the HF-PROM in the reported quantities.

Varying the number of directly computed HF snapshots revealed a non-monotonic relationship between initialization size and
prediction accuracy. Increasing this number from $4$ to $12$ reduced the snapshot-generation speedup from $7.8$ to $3.3$,
because the additional HF initialization cost exceeded the savings in enrichment. At a ROB dimension of $40$, initialization with
$10$ HF snapshots yielded the lowest mean aerodynamic errors among the tested MF-PROMs: $0.39\%$ for lift, $2.21\%$ for drag,
and $2.19\%$ for the lift-to-drag ratio. Increasing the initial HF snapshot count to $12$ increased these errors to $0.93\%$,
$6.29\%$, and $5.95\%$, respectively. The ranking depended on the ROB dimension and quantity of interest, so no initialization
size was uniformly most accurate.

This non-monotonic behavior is consistent with the configuration-dependent correction spaces and residual-based
greedy enrichment procedure. Changing the initial HF database can alter the corrected LF snapshots and subsequent enrichment
parameters; consequently, the final training databases and reduced spaces are not guaranteed to be nested. The observed errors
therefore do not support a general assumption that additional HF initialization improves the final MF-PROM. The preferred
initialization depends on the required accuracy, the quantities of interest, the ROB dimension, and the training cost.

The results also identify limitations of the present approach.  Beyond a ROB dimension of approximately $20$, the MF-PROM4 lift
error fluctuated near $1.1\%$, whereas the HF-PROM lift error decreased overall to $0.18\%$. At a selected test point, both
PROMs captured the overall shape of the shock-associated skin-friction feature but displaced its location. This behavior
is consistent with the difficulty of representing parameter-dependent transport of sharp features in a fixed
affine trial space. Future work will investigate trial spaces capable of transporting shocks and criteria for weighting
corrected LF snapshots relative to HF snapshots during basis construction. It will also develop indicators that assess in
advance whether a given LF model correlates sufficiently with the HF model for multifidelity construction to be beneficial.
Extensions to three-dimensional configurations, unsteady flows, and hyperreduced PROMs, together with quantification of the
uncertainty introduced by LF-derived training data, will further assess the applicability of the methodology.

\section*{Acknowledgments}
The authors acknowledge partial support from the Air Force Office of Scientific Research under Grants FA9550-22-1-0004 and 
FA9550-24-1-0327, and from the Office of Naval Research under Grant N00014-23-1-2877.

\bibliography{references}

@article{kennedy2000predicting,
  title={Predicting the output from a complex computer code when fast approximations are available},
  author={Kennedy, Marc C and O'Hagan, Anthony},
  journal={Biometrika},
  volume={87},
  number={1},
  pages={1--13},
  year={2000},
  publisher={Oxford University Press}
}

@article{aretz2026multifidelity,
  author  = {Aretz, Nicole and Willcox, Karen},
  title   = {Multifidelity proper orthogonal decomposition},
  journal = {arXiv preprint arXiv:2605.29213},
  year    = {2026},
  doi     = {10.48550/arXiv.2605.29213}
}

@article{behzad2018multilevel,
  title={Multilevel algorithm for obtaining the proper orthogonal decomposition},
  author={Behzad, Fariduddin and Helenbrook, Brian T and Ahmadi, Goodarz},
  journal={AIAA Journal},
  volume={56},
  number={11},
  pages={4423--4436},
  year={2018},
  publisher={American Institute of Aeronautics and Astronautics}
}

@article{conti2024multi,
  title={Multi-fidelity reduced-order surrogate modelling},
  author={Conti, Paolo and Guo, Mengwu and Manzoni, Andrea and Frangi, Attilio and Brunton, Steven L and Kutz, J. Nathan},
  journal={Proceedings of the Royal Society A: Mathematical, Physical and Engineering Sciences},
  volume={480},
  number={2283},
  pages={20230655},
  year={2024},
  publisher={The Royal Society}
}

@article{hampton2018practical,
  title={Practical error bounds for a non-intrusive bi-fidelity approach to parametric/stochastic model reduction},
  author={Hampton, Jerrad and Fairbanks, Hillary R and Narayan, Akil and Doostan, Alireza},
  journal={Journal of Computational Physics},
  volume={368},
  pages={315--332},
  year={2018},
  publisher={Elsevier}
}

@article{binev2011convergence,
  title={Convergence rates for greedy algorithms in reduced basis methods},
  author={Binev, Peter and Cohen, Albert and Dahmen, Wolfgang and DeVore, Ronald and Petrova, Guergana and Wojtaszczyk, Przemyslaw},
  journal={SIAM Journal on Mathematical Analysis},
  volume={43},
  number={3},
  pages={1457--1472},
  year={2011},
  publisher={SIAM}
}

@article{chellappa2020adaptive,
  title={Adaptive basis construction and improved error estimation for parametric nonlinear dynamical systems},
  author={Chellappa, Sridhar and Feng, Lihong and Benner, Peter},
  journal={International Journal for Numerical Methods in Engineering},
  volume={121},
  number={23},
  pages={5320--5349},
  year={2020},
  publisher={Wiley Online Library}
}

@article{feng2023multi,
  title={Multi-fidelity error estimation accelerates greedy model reduction of complex dynamical systems},
  author={Feng, Lihong and Lombardi, Luigi and Antonini, Giulio and Benner, Peter},
  journal={International Journal for Numerical Methods in Engineering},
  volume={124},
  number={23},
  pages={5312--5333},
  year={2023},
  publisher={Wiley Online Library}
}

@article{grassle2018pod,
  title={{POD} reduced-order modeling for evolution equations utilizing arbitrary finite element discretizations},
  author={Gr{\"a}{\ss}le, Carmen and Hinze, Michael},
  journal={Advances in Computational Mathematics},
  volume={44},
  number={6},
  pages={1941--1978},
  year={2018},
  publisher={Springer}
}

@article{grassle2019pod,
  title={{POD} model order reduction with space-adapted snapshots for incompressible flows},
  author={Gr{\"a}{\ss}le, Carmen and Hinze, Michael and Lang, Jens and Ullmann, Sebastian},
  journal={Advances in Computational Mathematics},
  volume={45},
  number={5},
  pages={2401--2428},
  year={2019},
  publisher={Springer}
}

@article{ullmann2016pod,
  title={{POD-Galerkin} reduced-order modeling with adaptive finite element snapshots},
  author={Ullmann, Sebastian and Rotkvic, Marko and Lang, Jens},
  journal={Journal of Computational Physics},
  volume={325},
  pages={244--258},
  year={2016},
  publisher={Elsevier}
}

@article{carlberg2011efficient,
  title={Efficient non-linear model reduction via a least-squares {Petrov--Galerkin} projection and compressive tensor approximations},
  author={Carlberg, Kevin and Bou-Mosleh, Charbel and Farhat, Charbel},
  journal={International Journal for Numerical Methods in Engineering},
  volume={86},
  number={2},
  pages={155--181},
  year={2011},
  publisher={Wiley Online Library}
}

@article{carlberg2013gnat,
  title={The {GNAT} method for nonlinear model reduction: Effective implementation and application to computational fluid dynamics and turbulent flows},
  author={Carlberg, Kevin and Farhat, Charbel and Cortial, Julien and Amsallem, David},
  journal={Journal of Computational Physics},
  volume={242},
  pages={623--647},
  year={2013},
  publisher={Elsevier}
}

@article{farrell2009conservative,
  title={Conservative interpolation between unstructured meshes via supermesh construction},
  author={Farrell, Patrick E and Piggott, Matthew D and Pain, Christopher C and Gorman, Gerard J and Wilson, Cian R},
  journal={Computer Methods in Applied Mechanics and Engineering},
  volume={198},
  number={33-36},
  pages={2632--2642},
  year={2009},
  publisher={Elsevier}
}

@article{reiss2018shifted,
  title={The shifted proper orthogonal decomposition: A mode decomposition for multiple transport phenomena},
  author={Reiss, Julius and Schulze, Philipp and Sesterhenn, J{\"o}rn and Mehrmann, Volker},
  journal={SIAM Journal on Scientific Computing},
  volume={40},
  number={3},
  pages={A1322--A1344},
  year={2018},
  publisher={SIAM}
}
\end{document}